 \documentclass[twocolumn,aps,prc,longbibliography,superscriptaddress,nofootinbib,showpacs,floatfix]{revtex4-1}
\usepackage{graphicx}   
\usepackage{xspace}     

\newcommand{\sqsn}{\mbox{$\sqrt{s_{_{NN}}}$}\xspace}
\newcommand {\pp}{\mbox{$p$$+$$p$}\xspace}
\newcommand{\dAu}{\mbox{$d$$+$Au}\xspace}

\newcommand{\ee}{\mbox{$e^+e^-$}\xspace}
\newcommand{\epem}{\mbox{$e^+e^-$}\xspace}

\newcommand{\gm}{\mbox{$(g-2)_\mu$}\xspace}

\begin{document}

\title{Search for dark photons from neutral meson decays in $p$$+$$p$ and 
$d$$+$Au collisions at $\sqrt{s_{_{NN}}}$=200 GeV}

\newcommand{\abilene}{Abilene Christian University, Abilene, Texas 79699, USA}
\newcommand{\acadsin}{Institute of Physics, Academia Sinica, Taipei 11529, Taiwan}
\newcommand{\augie}{Department of Physics, Augustana College, Sioux Falls, South Dakota 57197, USA}
\newcommand{\banaras}{Department of Physics, Banaras Hindu University, Varanasi 221005, India}
\newcommand{\barc}{Bhabha Atomic Research Centre, Bombay 400 085, India}
\newcommand{\baruch}{Baruch College, City University of New York, New York, New York, 10010 USA}
\newcommand{\bnlcoll}{Collider-Accelerator Department, Brookhaven National Laboratory, Upton, New York 11973-5000, USA}
\newcommand{\bnlphys}{Physics Department, Brookhaven National Laboratory, Upton, New York 11973-5000, USA}
\newcommand{\caucr}{University of California - Riverside, Riverside, California 92521, USA}
\newcommand{\charlesczech}{Charles University, Ovocn\'{y} trh 5, Praha 1, 116 36, Prague, Czech Republic}
\newcommand{\chonbuk}{Chonbuk National University, Jeonju, 561-756, Korea}
\newcommand{\ciae}{Science and Technology on Nuclear Data Laboratory, China Institute of Atomic Energy, Beijing 102413, People's~Republic~of~China}
\newcommand{\cns}{Center for Nuclear Study, Graduate School of Science, University of Tokyo, 7-3-1 Hongo, Bunkyo, Tokyo 113-0033, Japan}
\newcommand{\colorado}{University of Colorado, Boulder, Colorado 80309, USA}
\newcommand{\columbia}{Columbia University, New York, New York 10027 and Nevis Laboratories, Irvington, New York 10533, USA}
\newcommand{\czechtech}{Czech Technical University, Zikova 4, 166 36 Prague 6, Czech Republic}
\newcommand{\dapnia}{Dapnia, CEA Saclay, F-91191, Gif-sur-Yvette, France}
\newcommand{\debrecen}{Debrecen University, H-4010 Debrecen, Egyetem t{\'e}r 1, Hungary}
\newcommand{\elte}{ELTE, E{\"o}tv{\"o}s Lor{\'a}nd University, H-1117 Budapest, P\'azmany P\'eter s\'et\'any 1/A, Hungary}
\newcommand{\ewha}{Ewha Womans University, Seoul 120-750, Korea}
\newcommand{\fit}{Florida Institute of Technology, Melbourne, Florida 32901, USA}
\newcommand{\fsu}{Florida State University, Tallahassee, Florida 32306, USA}
\newcommand{\gsu}{Georgia State University, Atlanta, Georgia 30303, USA}
\newcommand{\hanyang}{Hanyang University, Seoul 133-792, Korea}
\newcommand{\hiroshima}{Hiroshima University, Kagamiyama, Higashi-Hiroshima 739-8526, Japan}
\newcommand{\howard}{Department of Physics and Astronomy, Howard University, Washington, DC 20059, USA}
\newcommand{\ihepprot}{IHEP Protvino, State Research Center of Russian Federation, Institute for High Energy Physics, Protvino, 142281, Russia}
\newcommand{\illuiuc}{University of Illinois at Urbana-Champaign, Urbana, Illinois 61801, USA}
\newcommand{\inrras}{Institute for Nuclear Research of the Russian Academy of Sciences, prospekt 60-letiya Oktyabrya 7a, Moscow 117312, Russia}
\newcommand{\instpasczech}{Institute of Physics, Academy of Sciences of the Czech Republic, Na Slovance 2, 182 21 Prague 8, Czech Republic}
\newcommand{\isu}{Iowa State University, Ames, Iowa 50011, USA}
\newcommand{\jaea}{Advanced Science Research Center, Japan Atomic Energy Agency, 2-4 Shirakata Shirane, Tokai-mura, Naka-gun, Ibaraki-ken 319-1195, Japan}
\newcommand{\jinrdubna}{Joint Institute for Nuclear Research, 141980 Dubna, Moscow Region, Russia}
\newcommand{\jyvaskyla}{Helsinki Institute of Physics and University of Jyv{\"a}skyl{\"a}, P.O.Box 35, FI-40014 Jyv{\"a}skyl{\"a}, Finland}
\newcommand{\kek}{KEK, High Energy Accelerator Research Organization, Tsukuba, Ibaraki 305-0801, Japan}
\newcommand{\korea}{Korea University, Seoul, 136-701, Korea}
\newcommand{\kurchatov}{Russian Research Center ``Kurchatov Institute," Moscow, 123098 Russia}
\newcommand{\kyoto}{Kyoto University, Kyoto 606-8502, Japan}
\newcommand{\labllr}{Laboratoire Leprince-Ringuet, Ecole Polytechnique, CNRS-IN2P3, Route de Saclay, F-91128, Palaiseau, France}
\newcommand{\lahorelums}{Physics Department, Lahore University of Management Sciences, Lahore 54792, Pakistan}
\newcommand{\lawllnl}{Lawrence Livermore National Laboratory, Livermore, California 94550, USA}
\newcommand{\losalamos}{Los Alamos National Laboratory, Los Alamos, New Mexico 87545, USA}
\newcommand{\lpc}{LPC, Universit{\'e} Blaise Pascal, CNRS-IN2P3, Clermont-Fd, 63177 Aubiere Cedex, France}
\newcommand{\lund}{Department of Physics, Lund University, Box 118, SE-221 00 Lund, Sweden}
\newcommand{\maryland}{University of Maryland, College Park, Maryland 20742, USA}
\newcommand{\mass}{Department of Physics, University of Massachusetts, Amherst, Massachusetts 01003-9337, USA }
\newcommand{\michigan}{Department of Physics, University of Michigan, Ann Arbor, Michigan 48109-1040, USA}
\newcommand{\muenster}{Institut f\"ur Kernphysik, University of Muenster, D-48149 Muenster, Germany}
\newcommand{\muhlenberg}{Muhlenberg College, Allentown, Pennsylvania 18104-5586, USA}
\newcommand{\myongji}{Myongji University, Yongin, Kyonggido 449-728, Korea}
\newcommand{\nagasaki}{Nagasaki Institute of Applied Science, Nagasaki-shi, Nagasaki 851-0193, Japan}
\newcommand{\natmephi}{National Research Nuclear University, MEPhI, Moscow Engineering Physics Institute, Moscow, 115409, Russia}
\newcommand{\newmex}{University of New Mexico, Albuquerque, New Mexico 87131, USA }
\newcommand{\nmsu}{New Mexico State University, Las Cruces, New Mexico 88003, USA}
\newcommand{\ohio}{Department of Physics and Astronomy, Ohio University, Athens, Ohio 45701, USA}
\newcommand{\ornl}{Oak Ridge National Laboratory, Oak Ridge, Tennessee 37831, USA}
\newcommand{\orsay}{IPN-Orsay, Universite Paris Sud, CNRS-IN2P3, BP1, F-91406, Orsay, France}
\newcommand{\peking}{Peking University, Beijing 100871, People's~Republic~of~China}
\newcommand{\pnpi}{PNPI, Petersburg Nuclear Physics Institute, Gatchina, Leningrad Region, 188300, Russia}
\newcommand{\riken}{RIKEN Nishina Center for Accelerator-Based Science, Wako, Saitama 351-0198, Japan}
\newcommand{\rikjrbrc}{RIKEN BNL Research Center, Brookhaven National Laboratory, Upton, New York 11973-5000, USA}
\newcommand{\rikkyo}{Physics Department, Rikkyo University, 3-34-1 Nishi-Ikebukuro, Toshima, Tokyo 171-8501, Japan}
\newcommand{\saispbstu}{Saint Petersburg State Polytechnic University, St. Petersburg, 195251 Russia}
\newcommand{\saopaulo}{Universidade de S{\~a}o Paulo, Instituto de F\'{\i}sica, Caixa Postal 66318, S{\~a}o Paulo CEP05315-970, Brazil}
\newcommand{\seoulnat}{Department of Physics and Astronomy, Seoul National University, Seoul 151-742, Korea}
\newcommand{\stonybrkc}{Chemistry Department, Stony Brook University, SUNY, Stony Brook, New York 11794-3400, USA}
\newcommand{\stonycrkp}{Department of Physics and Astronomy, Stony Brook University, SUNY, Stony Brook, New York 11794-3800, USA}
\newcommand{\subatech}{SUBATECH (Ecole des Mines de Nantes, CNRS-IN2P3, Universit{\'e} de Nantes) BP 20722 - 44307, Nantes, France}
\newcommand{\tenn}{University of Tennessee, Knoxville, Tennessee 37996, USA}
\newcommand{\titech}{Department of Physics, Tokyo Institute of Technology, Oh-okayama, Meguro, Tokyo 152-8551, Japan}
\newcommand{\tsukuba}{Institute of Physics, University of Tsukuba, Tsukuba, Ibaraki 305, Japan}
\newcommand{\vandy}{Vanderbilt University, Nashville, Tennessee 37235, USA}
\newcommand{\waseda}{Waseda University, Advanced Research Institute for Science and Engineering, 17 Kikui-cho, Shinjuku-ku, Tokyo 162-0044, Japan}
\newcommand{\weizmann}{Weizmann Institute, Rehovot 76100, Israel}
\newcommand{\wigner}{Institute for Particle and Nuclear Physics, Wigner Research Centre for Physics, Hungarian Academy of Sciences (Wigner RCP, RMKI) H-1525 Budapest 114, POBox 49, Budapest, Hungary}
\newcommand{\yonsei}{Yonsei University, IPAP, Seoul 120-749, Korea}
\newcommand{\zagreb}{University of Zagreb, Faculty of Science, Department of Physics, Bijeni\v{c}ka 32, HR-10002 Zagreb, Croatia}
\affiliation{\abilene}
\affiliation{\acadsin}
\affiliation{\augie}
\affiliation{\banaras}
\affiliation{\barc}
\affiliation{\baruch}
\affiliation{\bnlcoll}
\affiliation{\bnlphys}
\affiliation{\caucr}
\affiliation{\charlesczech}
\affiliation{\chonbuk}
\affiliation{\ciae}
\affiliation{\cns}
\affiliation{\colorado}
\affiliation{\columbia}
\affiliation{\czechtech}
\affiliation{\dapnia}
\affiliation{\debrecen}
\affiliation{\elte}
\affiliation{\ewha}
\affiliation{\fit}
\affiliation{\fsu}
\affiliation{\gsu}
\affiliation{\hanyang}
\affiliation{\hiroshima}
\affiliation{\howard}
\affiliation{\ihepprot}
\affiliation{\illuiuc}
\affiliation{\inrras}
\affiliation{\instpasczech}
\affiliation{\isu}
\affiliation{\jaea}
\affiliation{\jinrdubna}
\affiliation{\jyvaskyla}
\affiliation{\kek}
\affiliation{\korea}
\affiliation{\kurchatov}
\affiliation{\kyoto}
\affiliation{\labllr}
\affiliation{\lahorelums}
\affiliation{\lawllnl}
\affiliation{\losalamos}
\affiliation{\lpc}
\affiliation{\lund}
\affiliation{\maryland}
\affiliation{\mass}
\affiliation{\michigan}
\affiliation{\muenster}
\affiliation{\muhlenberg}
\affiliation{\myongji}
\affiliation{\nagasaki}
\affiliation{\natmephi}
\affiliation{\newmex}
\affiliation{\nmsu}
\affiliation{\ohio}
\affiliation{\ornl}
\affiliation{\orsay}
\affiliation{\peking}
\affiliation{\pnpi}
\affiliation{\riken}
\affiliation{\rikjrbrc}
\affiliation{\rikkyo}
\affiliation{\saispbstu}
\affiliation{\saopaulo}
\affiliation{\seoulnat}
\affiliation{\stonybrkc}
\affiliation{\stonycrkp}
\affiliation{\subatech}
\affiliation{\tenn}
\affiliation{\titech}
\affiliation{\tsukuba}
\affiliation{\vandy}
\affiliation{\waseda}
\affiliation{\weizmann}
\affiliation{\wigner}
\affiliation{\yonsei}
\affiliation{\zagreb}
\author{A.~Adare} \affiliation{\colorado}
\author{S.~Afanasiev} \affiliation{\jinrdubna}
\author{C.~Aidala} \affiliation{\losalamos} \affiliation{\mass} \affiliation{\michigan}
\author{N.N.~Ajitanand} \affiliation{\stonybrkc}
\author{Y.~Akiba} \affiliation{\riken} \affiliation{\rikjrbrc}
\author{R.~Akimoto} \affiliation{\cns}
\author{H.~Al-Bataineh} \affiliation{\nmsu}
\author{H.~Al-Ta'ani} \affiliation{\nmsu}
\author{J.~Alexander} \affiliation{\stonybrkc}
\author{M.~Alfred} \affiliation{\howard}
% ...
\collaboration{PHENIX Collaboration} \noaffiliation

\date{}


\begin{abstract}


The standard model (SM) of particle physics is spectacularly successful, 
yet the measured value of the muon anomalous magnetic moment $(g-2)_\mu$ 
deviates from SM calculations by 3.6$\sigma$.  Several theoretical models 
attribute this to the existence of a ``dark photon,'' an additional U(1) 
gauge boson, which is weakly coupled to ordinary photons.  The PHENIX 
experiment at the Relativistic Heavy Ion Collider has searched for a dark 
photon, $U$, in $\pi^0,\eta \rightarrow \gamma e^+e^-$ decays and obtained 
upper limits of $\mathcal{O}(2\times10^{-6})$ on $U$-$\gamma$ mixing at 
90\% CL for the mass range $30<m_U<90$ MeV/$c^2$.  Combined with other 
experimental limits, the remaining region in the $U$-$\gamma$ mixing 
parameter space that can explain the $(g-2)_\mu$ deviation from its SM 
value is nearly completely excluded at the 90\% confidence level, with 
only a small region of $29<m_U<32$~MeV/$c^2$ remaining.

\end{abstract}

\pacs{25.75.Dw}  

\maketitle 


%
%


{\em Introduction.}
The standard model (SM) of particle physics provides unprecedented 
numerical accuracy for quantities such as the anomalous magnetic moment of 
the electron $(g-2)_e$, as 
well as predicting the existence of the vector bosons $W^\pm$ and $Z^0$ 
and the recently discovered Higgs boson. Hence, measurements which lie 
outside SM predictions warrant special scrutiny. One such result is the 
measured value of $(g-2)_\mu$ for the muon~\cite{Bennett:2006fi}, which 
deviates from SM calculations by 3.6$\sigma$~\cite{Agashe:2014kda}. An 
intriguing explanation for this discrepancy has been proposed 
by adding a ``dark'' gauge 
boson~\cite{Fayet:2007ua,Pospelov:2008zw,Endo:2012hp,Davoudiasl:2012ig}. 
While the possibility of a hidden U(1) gauge sector had been considered 
shortly after the advent of the Standard 
Model~\cite{Galison:1983pa,Holdom:1985ag}, it has recently gained more 
relevance, because it provides a simultaneous explanation of various 
beyond-the-standard-model phenomena in addition to $(g-2)_\mu$. These 
include, for example, the discrepancy between the world's data on proton 
charge radius~\cite{Mohr:2008fa} and that obtained by the Lamb shift in 
muonic hydrogen~\cite{Pohl:2010zza,Antognini:1900ns}, and the positron 
excess in cosmic rays observed by ATIC~\cite{Chang:2008aa}, 
PAMELA~\cite{Adriani:2008zr} and AMS-II~\cite{Aguilar:2013qda} by 
providing a new mechanism for the decay of dark 
matter~\cite{ArkaniHamed:2008qn,TuckerSmith:2010ra}.

While a variety of mechanisms can be introduced to parameterize dark 
sector physics, a simple formulation postulates a ``dark photon'' of mass 
$m_U$ which mixes with QED photons via a ``kinetic coupling'' term in the 
Lagrangian~\cite{Galison:1983pa,Holdom:1985ag,Jaeckel:2013ija,Essig:2013lka}
\begin{equation}
 {\cal L}_{\rm mix} = -\frac{\varepsilon}{2}F_{\mu\nu}^{\rm QED}F^{\mu\nu}_{\rm dark},
\label{eq:Lag_mix}
\end{equation}
where $\varepsilon$ parametrizes the mixing strength. Dark photons can then 
mix with QED photons through all processes that involve QED photons, with 
an effective strength $\alpha_U = \varepsilon^2 \alpha_{EM}$. If the dark 
photon mass exceeds twice the electron mass, it can decay into an \epem 
pair, and in the minimal version of the model, this is its dominant decay 
mode in the interval $ 2m_e < m_U < 2m_\mu$. To date, a wide range of 
searches~\cite{Essig:2013lka} have excluded most of the $[m_U,\varepsilon]$ 
parameter space that could explain the deviation of $(g-2)_\mu$ from its 
SM value. In this work, we report on new limits that exclude at the 90\% 
confidence level essentially all of the remaining allowed parameter space, 
thereby rendering the dark photon an unlikely candidate to resolve the 
discrepancy of $(g-2)_\mu$ with the Standard Model.


{\em Searching for $\pi^0,\eta\rightarrow\gamma U, U\rightarrow e^+e^-$.} 
We search for possible decays of $\pi^0,\eta\rightarrow\gamma U, 
U\rightarrow e^+e^-$ by examining the invariant mass $m_{ee}$ of \ee pairs 
in a large sample of Dalitz decays, $\pi^0,\eta\rightarrow\gamma e^+e^-$ 
for $30<m_U<90~$MeV/$c^2$ in the dark photon parameter space, where the 
possibility of disentangling the $(g-2)_{\mu}$ anomaly by the dark photon 
survives at the 90\% confidence level. The invariant yield of virtual 
photons from the Dalitz decays of $\pi^0, \eta$ is given by the Kroll-Wada 
equation~\cite{Kroll:1955zu}:

\begin{equation}
\left(\frac{dN_{ee}}{dm_{ee}}\right)_{\gamma e^+e^-} = 
N_{2\gamma} \frac{4\alpha_{EM}}{3\pi} \frac{1}{m_{ee}} KW_{\pi^0,\eta}(m_{ee}) |F(m_{ee}^2)|^2, 
\label{eq:KW_ee}
\end{equation}
where
\begin{equation}
KW_{\pi^0,\eta}(m_{ee}) = \sqrt{1-\frac{4m_e^2}{m_{ee}^2}}
\left(1+\frac{2m_e^2}{m_{ee}^2}\right) \left(1-\frac{m_{ee}^{2}}{m_{\pi^0,\eta}^2}\right)^3,
\label{eq:KW_common}
\end{equation}
$N_{2\gamma}$ is the invariant yield of $2\gamma$ decays of $\pi^0,\eta$, 
$\alpha_{EM}$ is the fine structure constant, and $m_e, m_{\pi^0,\eta}$ 
are masses for the electron, $\pi^0$ and $\eta$, respectively.  The 
deviation of the transition form factor $F(q^2)$ from unity is 0.0157 even 
at $m_{ee}=90~$MeV/$c^2$ from the parameterization of 
$F(q^2)=(1-q^2/\Lambda^2)^{-1}$ with 
$\Lambda=0.72~$GeV~\cite{Dzhelyadin:1980kh}. Therefore, the variation of 
$F(q^2)$ is small enough in the mass range of interest to set $F(q^2)=1$ 
in the calculation. The weak coupling of the dark photon to the QED photon 
implies that the natural width of the dark photon is very narrow, and as a 
result the expected line shape of the dark photon is set by the mass 
resolution, $\sigma$, of the detector
\begin{equation}
\left(\frac{dN_{ee}}{dm_{ee}}\right)_{\gamma U} =
N_{2\gamma} \frac{2\varepsilon^2}{\sqrt{2\pi}\sigma}e^{\frac{-(m_{ee}-m_{U})^2}{2\sigma^2}}
  KW_{\pi^0,\eta}(m_{ee}).
\label{eq:dp_ee}
\end{equation}
From the peak height ratio, 
\begin{equation}
R(m_U)= (dN_{ee}/dm_{ee})_{\pi^0,\eta\rightarrow\gamma
  U}/(dN_{ee}/dm_{ee})_{\pi^0,\eta\rightarrow\gamma e^+e^-}, 
\label{eq:ph_ratio}
\end{equation}
the dark photon mixing parameter can then be determined as:
\begin{equation}
  \varepsilon^2 = \frac{2\alpha_{EM}}{3\pi} \frac{\sigma}{m_U} \sqrt{2\pi} R(m_U).
\label{eq:mix_par}
\end{equation}
Note that in this approach the efficiencies for detection of \ee pairs from 
Dalitz decays and from dark photons cancel in the ratio $R(m_U)$.

The analysis presented here is based on a precise measurement of virtual 
photons from $\pi^0$ and $\eta$ Dalitz decays~\cite{Adare:2012vn} across 
three PHENIX data sets at a collision energy of $\sqsn = 200$~GeV with an 
integrated luminosity of 4.8~pb$^{-1}$ of \pp collected in 2006, 
82.3~nb$^{-1}$ of \dAu collected in 2008, and 6.0~pb$^{-1}$ of \pp 
collected in 2009. Here, the \dAu statistics corresponds to 
$2\times197\times82.3$~nb$^{-1}$ = 32.4~pb$^{-1}$ of nucleon-nucleon 
collisions. All three data sets include an electron triggered sample, and 
the single electron trigger threshold for the \dAu run was higher than 
that for the \pp runs. A hadron blind detector 
(HBD)~\cite{Anderson:2011jw}, was installed in the experiment around the 
primary collision point prior to the 2009 data taking period.  The 
additional material of the HBD resulted in a corresponding increase in the 
external photon conversion rate.  The experiment was also operated with a 
reduced magnetic field integral during the period of HBD data taking.  
These effects substantially alter the shape of the 2009 \epem mass 
spectrum below 35~MeV/$c^2$ relative to the spectra from 2006 and 2008.
Therefore, we restrict the 2009 analysis to the mass region
above 40~MeV/$c^2$ to avoid the edge effect at parameterization of the 
Dalitz contribution.

The PHENIX apparatus~\cite{Adcox:2003zm} was designed with only 0.39\% of 
a radiation length ($X_0$) in front of the tracking detectors. It generates a 
small rate of conversions in the experimental aperture and provides 
excellent momentum resolution and electron identification. The HBD brought 
an additional material budget of $2.4\% \times X_0~$ for the 2009 run. The 
tracking system comprises drift wire and pad chambers with a momentum 
resolution of $\delta p/p=1\% \oplus 1.1\% \times p$~[GeV/$c$]. Charged 
tracks with momenta above 0.2~GeV/$c$ and pseudorapidity $|\eta| < 0.35$ 
fall within the PHENIX acceptance.  Electron identification requires hits 
in a Ring Imaging \v{C}erenkov detector and energy-momentum matching in an 
electromagnetic calorimeter with an energy resolution of $\delta 
E/E<10\%/\sqrt{E~\mbox{[GeV]}}$.

\begin{figure*}[htb]
\includegraphics[width=0.998\linewidth]{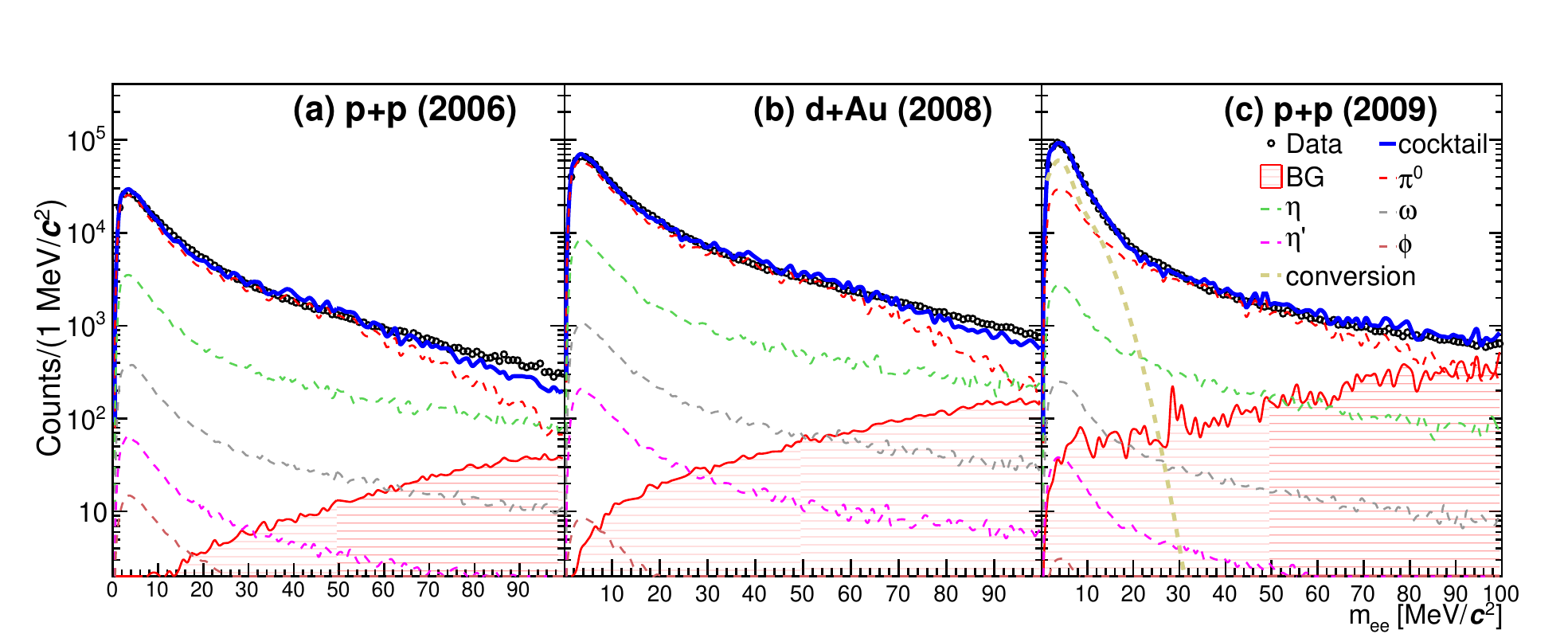}
\caption{\label{fig:mass_spec} 
(Color online)
The raw spectra of \epem pairs for the 2006 \pp, 
2008 \dAu and 2009 \pp data sets.  The contributions of 
various background components to the measured invariant 
mass spectra are shown.  The 2009 \pp data has a 
significant contribution to the conversion background 
coming from the material of the HBD which is not present 
in the 2006 and 2008 data sets.
}
\end{figure*}

All combinations of electrons and positrons in an event are taken as pairs 
for the analysis.  The contributions due to random combinations, 
correlated fake pairs from double Dalitz decays ($\pi^0, \eta \rightarrow 
e^+e^-e^+e^-$) and jet-induced correlations are evaluated using like-sign 
pairs.  After scaling by the number of nucleon-nucleon collisions, the 
correlated backgrounds in \pp and \dAu are very similar, indicating these 
background contributions are well understood.  Pairs stemming from photon 
conversions in the material of the detector are removed by a cut on their 
characteristic angular orientation with respect to the magnetic 
field~\cite{Adare:2009qk}.  For the 2009 \pp data, conversion pairs are 
rejected by a cut on the cluster size in the HBD, which depends on the 
pair opening angle~\cite{Adare:2012vv}, because the lower magnetic field 
of the 2009 run reduces the rejection power of the angular orientation 
cut.  Conversions in the HBD readout plane were removed by an analysis 
technique of mass reconstruction assuming electrons come from the HBD 
readout plane~\cite{PPG162:arXiv}. 
In the 2009 dataset we consider pairs with an 
invariant mass above 40~MeV/$c^2$, where the contribution of conversion 
pairs becomes negligible.  Excluding these nonhadronic background pairs, we 
obtained 67k, 167k and 75k \ee pairs for 2006 \pp, 2008 \dAu, and 2009 
\pp, respectively in the mass range $30<m_{ee}<90$~MeV/$c^2$, where most 
pairs originate from $\pi^0, \eta$ Dalitz decays.  Contributions to the 
electron pair spectrum are estimated by a GEANT3 based detector simulation 
using the measured invariant yields for hadrons as input.  Effects such as 
the single electron trigger efficiency and inactive areas in the detector 
are taken into account.  Figure~\ref{fig:mass_spec} shows the raw spectra 
of \epem pairs with the hadronic decay and background contributions for 
the 2006 \pp, 2008 $d$$+$Au and 2009 \pp data sets.

If the expected dark photon invariant mass distribution follows a normal 
distribution, then the standard deviation is equal to the detector mass 
resolution, as already described.  This resolution is 
determined using a Monte Carlo procedure based on a GEANT3 description of 
the experimental apparatus.  Spectra of dark photons with a flat 
distribution in transverse momentum for $p_T < 5$~GeV/$c$, covering the 
full azimuth, with rapidity $|y| < 0.5$, and with an initial vertex within 
35~cm of the nominal vertex position are generated and forced to decay as 
$U\rightarrow\epem$. Dark photon masses from 20--90~MeV/$c^2$ were 
investigated, with 20 million decays generated at each mass hypothesis.  
The reconstructed \epem pairs were then weighted according to their pair 
$p_T$ to follow the experimental \ee pair spectrum after background 
subtraction. The \ee invariant mass resolution for the PHENIX detector in 
$30<m_{ee}<90~$MeV/$c^2$ is $\sigma = 3.1~$MeV/$c^2$ with a 3\% 
uncertainty. The calculated mass resolution is also confirmed with the 
data via a shape matching of the $\pi^0$ Dalitz peak around 5~MeV/$c^2$.

To establish a limit on the dark photon yield, we first describe the shape 
of the background-subtracted \epem spectrum with a physics motivated curve 
composed of the Kroll-Wada formula for virtual photon yield from both the 
$\pi^0$ and the $\eta$ multiplied by a 4$^{\rm th}$-order Chebychev 
polynomial $T_4(x)$ to allow for slight deviations due to various detector 
effects:
\begin{equation}
  \label{eq:fit_KW}
  f(m_{ee}) = \frac{1}{m_{ee}} \times
  \left[\left(1-\frac{m_{ee}^2}{m_{\pi^0}^2}\right)^3 +
  r_{\eta/\pi^0}\times\left(1-\frac{m_{ee}^2}{m_{\eta}^2}\right)^3 \right] 
  \times  T_4(m_{ee}).
\end{equation}
 
The $\eta/\pi^0$ ratio, $r_{\eta/\pi^0}$, is fixed at 0.17, a value 
determined using a realistic ``cocktail'' of hadronic decays filtered 
through a model of the detector acceptance.  The $\omega/\pi^0$ ratio is 
fixed at 0.03. The shapes of the \ee mass spectra from $\eta$ and $\omega$ 
decays are indistinguishable for $m_{ee} < 100~$MeV/$c^2$, and their 
combined yield relative to the $\pi^0$, $0.17+0.03=0.20$, is taken as the 
effective $\eta/\pi^0$ ratio for the analysis.

We divide the full mass ranges of $25<m_{ee}<95$~MeV/$c^2$ and 
$35<m_{ee}<95$~MeV/$c^2$ into lower and higher mass ranges after 
nonhadronic background subtraction, use Eq.~\ref{eq:fit_KW} to 
describe each portion, 
and demand continuity of the model at the mass where the two ranges abut.  A 
simultaneous fit to the three mass spectra, allowing each an independent 
normalization, results in a combined description of the Dalitz continuum.  
This procedure produces a lower reduced $\chi^2$ for the overall fit than 
using a single mass range for each dataset.  The break point dividing the 
lower and upper mass ranges was allowed to vary, with 61~MeV/$c^2$ giving 
the best reduced $\chi^2$.  
\begin{figure*}[htb]
\includegraphics[width=0.998\linewidth]{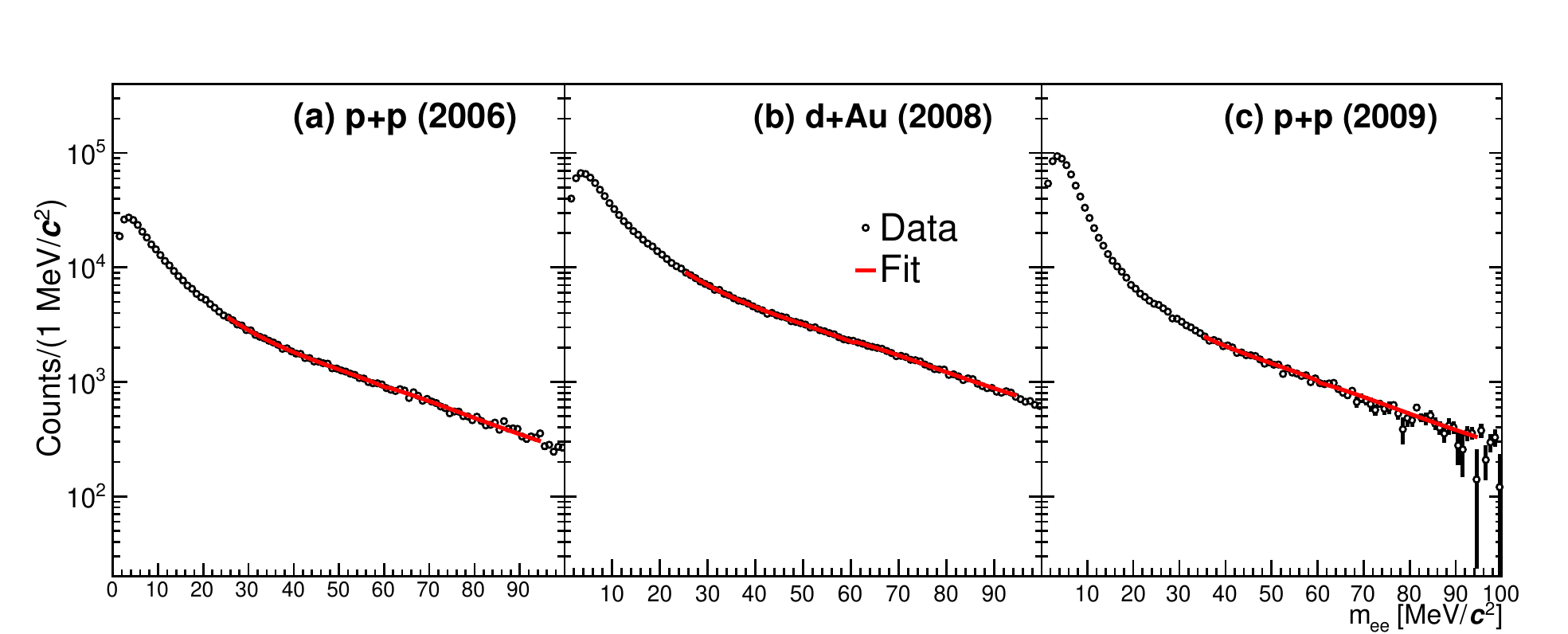}
\caption{\label{fig:fit_dalitz} 
(Color online)
The best fit to the three mass spectra with the physics motivated function 
describing the \ee distributions from hadron decays.
}
\end{figure*}
Figure~\ref{fig:fit_dalitz} shows the best fit result to the Dalitz decay 
contribution in each dataset after subtraction of unphysical background pairs.
The contribution of the fit procedure to the 
total uncertainty is explored by varying the break point above and below 
this preferred value until the reduced $\chi^2$ statistic rises by one and 
then taking the resulting 16\% effect on the experimental sensitivity as the 
systematic uncertainty due to the procedure.


\begin{figure*}[htb]
\includegraphics[width=0.998\linewidth]{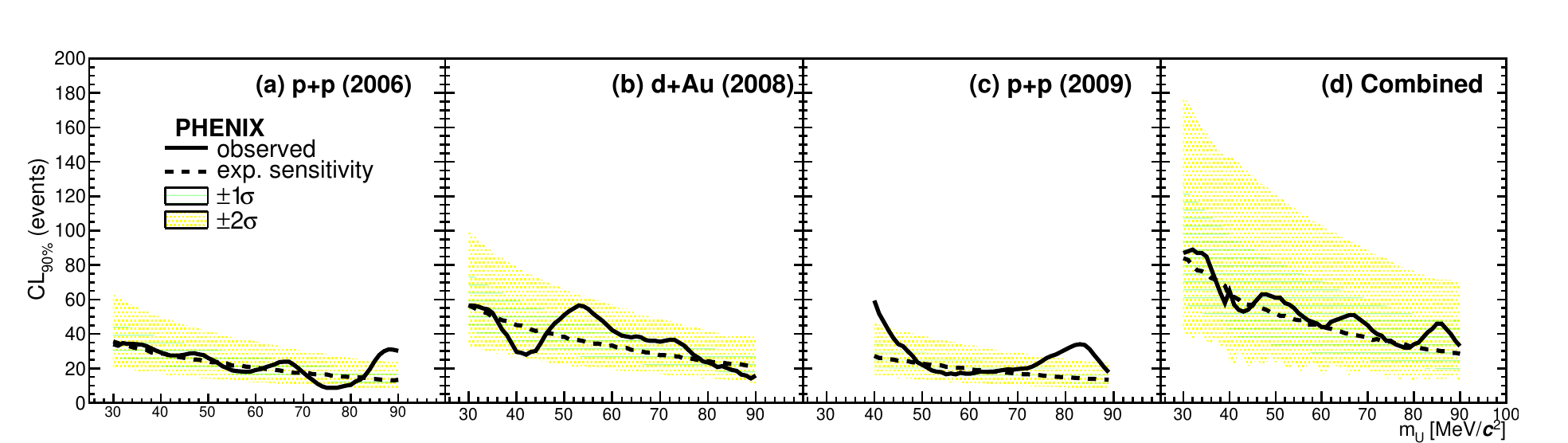}
\caption{\label{fig:exp_dphoton} 
(Color online)
The experimental sensitivity and observed limit on the number
of dark photon candidates as a function of the 
assumed dark photon mass.  The $\pm1\sigma$ and 
$\pm2\sigma$ bands of the combined statistical and 
systematic uncertainties around the experimental 
sensitivity are shown in green and yellow, respectively.
}
\end{figure*}

{\em Results.}
The fitted background describes the yield of \ee counts absent a dark 
photon signal.  We employ the CL$_s$ statistical 
approach~\cite{Read:2002hq} to determine a limit on the number of dark 
photon candidates, which is in line with the current practice of setting 
limits for a hypothetical particle.  This method has the effect of 
reducing the strength of the limit determination in the case of low (or 
no) signal strength, generally resulting in a conservative estimate of the 
CL.  We step through the full mass range with a 1~MeV/$c^2$ step repeatedly 
refitting the spectrum with the addition of a Gaussian of width equal to the mass 
resolution and centered at each mass hypothesis.  This determines the observed 
yield as a function of $m_U$, which may be greater or less than the experimental 
sensitivity at each mass, with a significance that is determined by the 
underlying probability distribution of the background, which is calculated 
by a likelihood ratio between the signal + background and background only 
hypotheses. 
The assumed background yield in any mass window will have uncertainties due to 
statistical fluctuations in the data used to determine the parameters 
describing the background by Eq.~\ref{eq:fit_KW} and from 
systematic uncertainties in alternative background shapes. We evaluated 
the variation in the experimental sensitivity due to fluctuations in these 
uncertainties in addition to the uncertainty in the \ee mass resolution. 
The observed value, the experimental sensitivity, and one- and two-standard 
deviation bands around the experimental sensitivity (shown as green and yellow 
bands) are all indicated on the plots for the different data sets as well 
as the combined result in Fig.~\ref{fig:exp_dphoton}.

\begin{figure}[htb]
\includegraphics[width=0.98\linewidth]{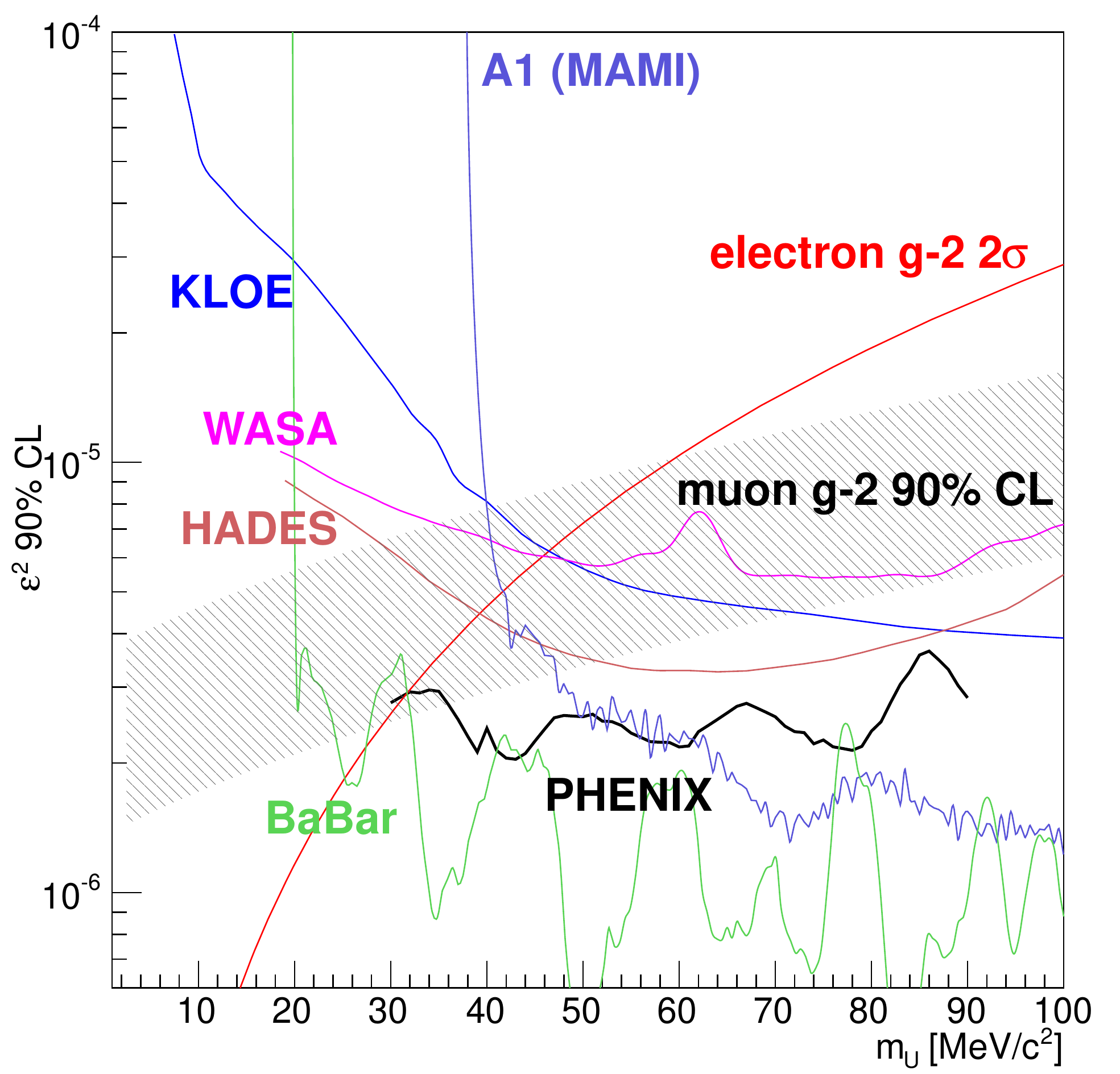}
\caption{\label{fig:dphotonlimit} 
(Color online)
A compilation of the limits on the $U$-$\gamma$ mixing parameter, showing 
the PHENIX results. Also shown are the limits at 90\% CL from 
WASA~\protect\cite{Adlarson:2013eza}, 
HADES~\protect\cite{Agakishiev:2013fwl}, 
KLOE~\protect\cite{Babusci:2012cr}, 
A1(MAMI)~\protect\cite{Merkel:2014avp}, and 
{\sc BaBar}~\protect\cite{Lees:2014xha} experiments and the band 
indicating the range of mass and coupling parameters favored 
by the \gm anomaly at 90\% CL.
Also shown is the $2\sigma$ upper limit obtained 
from $(g-2)_e$~\protect\cite{Davoudiasl:2014kua}.
}
\end{figure}

\begin{figure}[htb]
\includegraphics[width=0.98\linewidth]{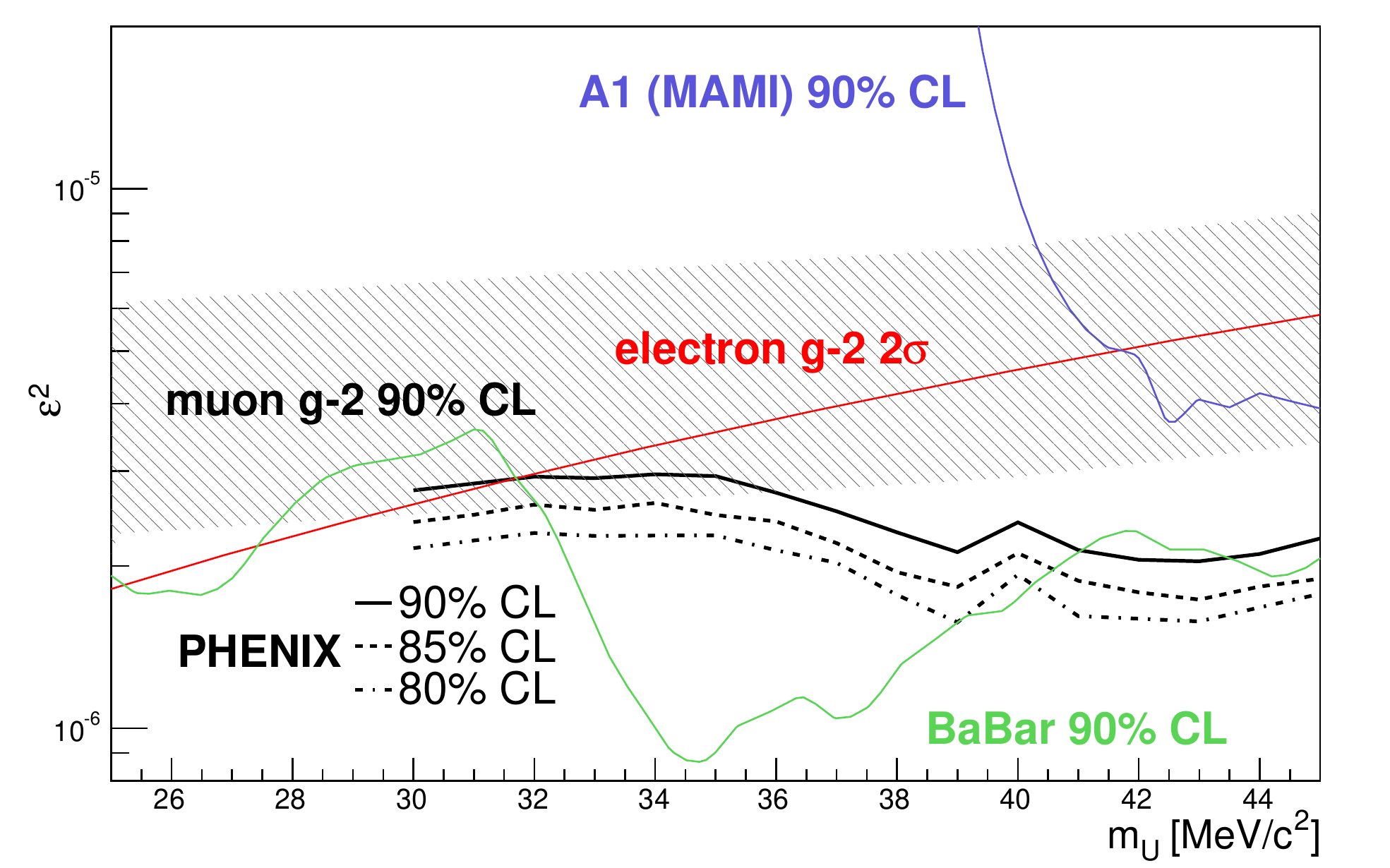}
\caption{\label{fig:dphotonprob} 
(Color online)
Limits on the $U$-$\gamma$ mixing parameters from PHENIX 
at different confidence levels, together with the 90\% CL limits 
from {\sc BaBar}~\cite{Lees:2014xha}, and A1(MAMI)~\cite{Merkel:2014avp}, 
the $2\sigma$ upper limit derived from $(g-2)_e$~\cite{Davoudiasl:2014kua} 
and the region favored by \gm.
}
\end{figure}

The $p$-value under the null hypothesis from the combined 
result is calculated considering only the statistical uncertainty and is 
always greater than 0.27 in the entire range $30<m_U<90~$MeV/$c^2$.  The 
minimum $p$-value is consistent with the background only hypothesis if the 
{\em look-elsewhere effect}~\cite{lookelse} is taken into account. 
Therefore the limit on the number of dark photon candidate events can be 
translated directly into a limit on the dark photon coupling parameter 
using the peak-height ratio, Eq.~\ref{eq:ph_ratio}.  
Figure~\ref{fig:dphotonlimit} shows the limit determined by PHENIX along 
with the 90\% confidence level (CL) limits from the WASA~\cite{Adlarson:2013eza}, 
HADES~\cite{Agakishiev:2013fwl}, KLOE~\cite{Babusci:2012cr}, 
A1(MAMI)~\cite{Merkel:2014avp} and {\sc BaBar}~\cite{Lees:2014xha} 
experiments and the 
$2\sigma$ upper limit theoretically calculated from 
$(g-2)_e$~\cite{Davoudiasl:2014kua}. The bands indicate the range of 
parameters which would allow the dark photon to explain the $(g-2)_\mu$ 
anomalies with the 90\% CL. The upward fluctuation apparent in the 2008 
\dAu data compensates for a downward fluctuation of similar scale in the 
2009 \pp data, leading to the slightly modulated limit of the combined 
result. The PHENIX results cover the mass range $30 < m_U < 90$~MeV/$c^2$, 
and over that range set a stricter limit than those of WASA, HADES or 
KLOE, and complement the A1(MAMI) results for their less sensitive region below 
50~MeV/$c^2$. The PHENIX limits exclude the values of the coupling favored 
by the $(g-2)_\mu$ anomaly above $m_U>36~$MeV/$c^2$. Recently, {\sc BaBar} 
reported stricter limits from a search of the reaction 
$e^+e^-\rightarrow\gamma U, U\rightarrow l^+l^-$,
excluding values of the preferred $(g-2)_\mu$ region for $m_U>32~$MeV/$c^2$, 
and covering a mass range up to 10.2~GeV/$c^2$. As a result, 
nearly all the available parameter space which would allow the dark photon 
to explain the $(g-2)_\mu$ results are ruled out at the 90\% CL by 
independent experiments.  Figure~\ref{fig:dphotonprob} shows the PHENIX 
limits in the dark photon parameter space with different confidence 
levels, focusing on the small remaining parameter space for 
$30<m_U<32~$MeV/$c^2$. The entire parameter space to explain the \gm 
anomaly by the dark photon can be excluded at the 85\% CL by the PHENIX 
data alone. The level of the compatibility between our data and the 
coupling strength favored for the $(g-2)_\mu$ anomaly is 10\% with a 
statistical test~\cite{Maltoni:2003cu}.

{\em Conclusions.} In summary, the PHENIX results set limits for the 
coupling of a dark photon to the QED photon over the mass range $30 < m_U 
< 90$~MeV/$c^2$, improving upon the recent results of the KLOE, WASA, 
HADES, and A1 experiments.  Combining with the {\sc BaBar} results, the 
dark photon is ruled out at the 90\% CL as an explanation for the 
$(g-2)_\mu$ anomaly for $m_U >32$~MeV/$c^2$, leaving only a small 
remaining part of parameter space in the region $29<m_U<32$~MeV/$c^2$. 
The probability that the theoretically predicted coupling strength 
required to explain the $(g-2)_\mu$ anomaly is compatible with the PHENIX 
results is only 10\%. Future analyses by PHENIX would be able to provide 
even more stringent limits due to both increased data sets and improved 
detector technology that allow measurement of displaced vertices. As the 
coupling to the dark photon gets weaker, the distance traveled by the 
dark photon before decaying into \epem grows 
longer~\cite{Bjorken:2009mm}.  The high statistics dataset taken after 
the recently commissioned PHENIX silicon vertex detector was installed in 
2011 is being analyzed to look for such weakly coupled dark photons to 
provide limits even more restrictive than those reported here.



{\em Acknowledgments.}
We thank the staff of the Collider-Accelerator and Physics
Departments at Brookhaven National Laboratory and the staff of
the other PHENIX participating institutions for their vital
contributions.  
We also thank William Marciano and Hye-Sung Lee for useful discussions and
theoretical calculations, and we thank the WASA, HADES and {\sc BaBar}
collaborations for useful interactions.
We acknowledge support from the 
Office of Nuclear Physics in the
Office of Science of the Department of Energy, 
the National Science Foundation, 
a sponsored research grant from Renaissance Technologies LLC,
Abilene Christian University Research Council, 
Research Foundation of SUNY, and 
Dean of the College of Arts and Sciences, Vanderbilt University 
(U.S.A),
Ministry of Education, Culture, Sports, Science, and Technology
and the Japan Society for the Promotion of Science (Japan),
Conselho Nacional de Desenvolvimento Cient\'{\i}fico e
Tecnol{\'o}gico and Funda\c c{\~a}o de Amparo {\`a} Pesquisa do
Estado de S{\~a}o Paulo (Brazil),
Natural Science Foundation of China (P.~R.~China),
Ministry of Science, Education, and Sports (Croatia),
Ministry of Education, Youth and Sports (Czech Republic),
Centre National de la Recherche Scientifique, Commissariat
{\`a} l'{\'E}nergie Atomique, and Institut National de Physique
Nucl{\'e}aire et de Physique des Particules (France),
Bundesministerium f\"ur Bildung und Forschung, Deutscher
Akademischer Austausch Dienst, and Alexander von Humboldt Stiftung (Germany),
OTKA NK 101 428 grant and the Ch. Simonyi Fund (Hungary),
Department of Atomic Energy and Department of Science and Technology (India), 
Israel Science Foundation (Israel), 
Basic Science Research Program through NRF of the Ministry of Education (Korea),
Physics Department, Lahore University of Management Sciences (Pakistan),
Ministry of Education and Science, Russian Academy of Sciences,
Federal Agency of Atomic Energy (Russia),
VR and Wallenberg Foundation (Sweden), 
the U.S. Civilian Research and Development Foundation for the
Independent States of the Former Soviet Union, 
the Hungarian American Enterprise Scholarship Fund,
and the US-Israel Binational Science Foundation.



\begin{thebibliography}{36}%
\makeatletter
\providecommand \@ifxundefined [1]{%
 \@ifx{#1\undefined}
}%
\providecommand \@ifnum [1]{%
 \ifnum #1\expandafter \@firstoftwo
 \else \expandafter \@secondoftwo
 \fi
}%
\providecommand \@ifx [1]{%
 \ifx #1\expandafter \@firstoftwo
 \else \expandafter \@secondoftwo
 \fi
}%
\providecommand \natexlab [1]{#1}%
\providecommand \enquote  [1]{``#1''}%
\providecommand \bibnamefont  [1]{#1}%
\providecommand \bibfnamefont [1]{#1}%
\providecommand \citenamefont [1]{#1}%
\providecommand \href@noop [0]{\@secondoftwo}%
\providecommand \href [0]{\begingroup \@sanitize@url \@href}%
\providecommand \@href[1]{\@@startlink{#1}\@@href}%
\providecommand \@@href[1]{\endgroup#1\@@endlink}%
\providecommand \@sanitize@url [0]{\catcode `\\12\catcode `\$12\catcode
  `\&12\catcode `\#12\catcode `\^12\catcode `\_12\catcode `\%12\relax}%
\providecommand \@@startlink[1]{}%
\providecommand \@@endlink[0]{}%
\providecommand \url  [0]{\begingroup\@sanitize@url \@url }%
\providecommand \@url [1]{\endgroup\@href {#1}{\urlprefix }}%
\providecommand \urlprefix  [0]{URL }%
\providecommand \Eprint [0]{\href }%
\providecommand \doibase [0]{http://dx.doi.org/}%
\providecommand \selectlanguage [0]{\@gobble}%
\providecommand \bibinfo  [0]{\@secondoftwo}%
\providecommand \bibfield  [0]{\@secondoftwo}%
\providecommand \translation [1]{[#1]}%
\providecommand \BibitemOpen [0]{}%
\providecommand \bibitemStop [0]{}%
\providecommand \bibitemNoStop [0]{.\EOS\space}%
\providecommand \EOS [0]{\spacefactor3000\relax}%
\providecommand \BibitemShut  [1]{\csname bibitem#1\endcsname}%
\let\auto@bib@innerbib\@empty
\bibitem [{\citenamefont {Bennett}\ \emph {et~al.}(2006)\citenamefont {Bennett}
  \emph {et~al.}}]{Bennett:2006fi}%
  \BibitemOpen
  \bibfield  {author} {\bibinfo {author} {\bibfnamefont {G.W.}\ \bibnamefont
  {Bennett}} \emph {et~al.} (\bibinfo {collaboration} {Muon G-2
  Collaboration}),\ }\bibfield  {title} {\enquote {\bibinfo {title} {{Final
  Report of the Muon E821 Anomalous Magnetic Moment Measurement at BNL}},}\
  }\href {\doibase 10.1103/PhysRevD.73.072003} {\bibfield  {journal} {\bibinfo
  {journal} {Phys. Rev. D}\ }\textbf {\bibinfo {volume} {73}},\ \bibinfo
  {pages} {072003} (\bibinfo {year} {2006})}\BibitemShut {NoStop}%
\bibitem [{\citenamefont {Olive}\ \emph {et~al.}(2014)\citenamefont {Olive}
  \emph {et~al.}}]{Agashe:2014kda}%
  \BibitemOpen
  \bibfield  {author} {\bibinfo {author} {\bibfnamefont {K.~A.}\ \bibnamefont
  {Olive}} \emph {et~al.} (\bibinfo {collaboration} {Particle Data Group}),\
  }\bibfield  {title} {\enquote {\bibinfo {title} {{Review of Particle
  Physics}},}\ }\href {\doibase 10.1088/1674-1137/38/9/090001} {\bibfield
  {journal} {\bibinfo  {journal} {Chin. Phys. C}\ }\textbf {\bibinfo {volume}
  {38}},\ \bibinfo {pages} {090001} (\bibinfo {year} {2014})}\BibitemShut
  {NoStop}%
\bibitem [{\citenamefont {Fayet}(2007)}]{Fayet:2007ua}%
  \BibitemOpen
  \bibfield  {author} {\bibinfo {author} {\bibfnamefont {Pierre}\ \bibnamefont
  {Fayet}},\ }\bibfield  {title} {\enquote {\bibinfo {title} {{U-boson
  production in e+ e- annihilations, psi and Upsilon decays, and Light Dark
  Matter}},}\ }\href {\doibase 10.1103/PhysRevD.75.115017} {\bibfield
  {journal} {\bibinfo  {journal} {Phys. Rev. D}\ }\textbf {\bibinfo {volume}
  {75}},\ \bibinfo {pages} {115017} (\bibinfo {year} {2007})}\BibitemShut
  {NoStop}%
\bibitem [{\citenamefont {Pospelov}(2009)}]{Pospelov:2008zw}%
  \BibitemOpen
  \bibfield  {author} {\bibinfo {author} {\bibfnamefont {Maxim}\ \bibnamefont
  {Pospelov}},\ }\bibfield  {title} {\enquote {\bibinfo {title} {{Secluded U(1)
  below the weak scale}},}\ }\href {\doibase 10.1103/PhysRevD.80.095002}
  {\bibfield  {journal} {\bibinfo  {journal} {Phys. Rev. D}\ }\textbf {\bibinfo
  {volume} {80}},\ \bibinfo {pages} {095002} (\bibinfo {year}
  {2009})}\BibitemShut {NoStop}%
\bibitem [{\citenamefont {Endo}\ \emph {et~al.}(2012)\citenamefont {Endo},
  \citenamefont {Hamaguchi},\ and\ \citenamefont {Mishima}}]{Endo:2012hp}%
  \BibitemOpen
  \bibfield  {author} {\bibinfo {author} {\bibfnamefont {Motoi}\ \bibnamefont
  {Endo}}, \bibinfo {author} {\bibfnamefont {Koichi}\ \bibnamefont
  {Hamaguchi}}, \ and\ \bibinfo {author} {\bibfnamefont {Go}~\bibnamefont
  {Mishima}},\ }\bibfield  {title} {\enquote {\bibinfo {title} {{Constraints on
  Hidden Photon Models from Electron g-2 and Hydrogen Spectroscopy}},}\ }\href
  {\doibase 10.1103/PhysRevD.86.095029} {\bibfield  {journal} {\bibinfo
  {journal} {Phys. Rev. D}\ }\textbf {\bibinfo {volume} {86}},\ \bibinfo
  {pages} {095029} (\bibinfo {year} {2012})}\BibitemShut {NoStop}%
\bibitem [{\citenamefont {Davoudiasl}\ \emph {et~al.}(2012)\citenamefont
  {Davoudiasl}, \citenamefont {Lee},\ and\ \citenamefont
  {Marciano}}]{Davoudiasl:2012ig}%
  \BibitemOpen
  \bibfield  {author} {\bibinfo {author} {\bibfnamefont {Hooman}\ \bibnamefont
  {Davoudiasl}}, \bibinfo {author} {\bibfnamefont {Hye-Sung}\ \bibnamefont
  {Lee}}, \ and\ \bibinfo {author} {\bibfnamefont {William~J.}\ \bibnamefont
  {Marciano}},\ }\bibfield  {title} {\enquote {\bibinfo {title} {{Dark Side of
  Higgs Diphoton Decays and Muon g-2}},}\ }\href {\doibase
  10.1103/PhysRevD.86.095009} {\bibfield  {journal} {\bibinfo  {journal} {Phys.
  Rev. D}\ }\textbf {\bibinfo {volume} {86}},\ \bibinfo {pages} {095009}
  (\bibinfo {year} {2012})}\BibitemShut {NoStop}%
\bibitem [{\citenamefont {Galison}\ and\ \citenamefont
  {Manohar}(1984)}]{Galison:1983pa}%
  \BibitemOpen
  \bibfield  {author} {\bibinfo {author} {\bibfnamefont {Peter}\ \bibnamefont
  {Galison}}\ and\ \bibinfo {author} {\bibfnamefont {Aneesh}\ \bibnamefont
  {Manohar}},\ }\bibfield  {title} {\enquote {\bibinfo {title} {{Two Z's or not
  two Z's?}}}\ }\href {\doibase 10.1016/0370-2693(84)91161-4} {\bibfield
  {journal} {\bibinfo  {journal} {Phys. Lett. B}\ }\textbf {\bibinfo {volume}
  {136}},\ \bibinfo {pages} {279} (\bibinfo {year} {1984})}\BibitemShut
  {NoStop}%
\bibitem [{\citenamefont {Holdom}(1986)}]{Holdom:1985ag}%
  \BibitemOpen
  \bibfield  {author} {\bibinfo {author} {\bibfnamefont {Bob}\ \bibnamefont
  {Holdom}},\ }\bibfield  {title} {\enquote {\bibinfo {title} {{Two U(1)'s and
  Epsilon Charge Shifts}},}\ }\href {\doibase 10.1016/0370-2693(86)91377-8}
  {\bibfield  {journal} {\bibinfo  {journal} {Phys. Lett. B}\ }\textbf
  {\bibinfo {volume} {166}},\ \bibinfo {pages} {196} (\bibinfo {year}
  {1986})}\BibitemShut {NoStop}%
\bibitem [{\citenamefont {Mohr}\ \emph {et~al.}(2008)\citenamefont {Mohr},
  \citenamefont {Taylor},\ and\ \citenamefont {Newell}}]{Mohr:2008fa}%
  \BibitemOpen
  \bibfield  {author} {\bibinfo {author} {\bibfnamefont {Peter~J.}\
  \bibnamefont {Mohr}}, \bibinfo {author} {\bibfnamefont {Barry~N.}\
  \bibnamefont {Taylor}}, \ and\ \bibinfo {author} {\bibfnamefont {David~B.}\
  \bibnamefont {Newell}},\ }\bibfield  {title} {\enquote {\bibinfo {title}
  {{CODATA Recommended Values of the Fundamental Physical Constants: 2006}},}\
  }\href {\doibase 10.1103/RevModPhys.80.633} {\bibfield  {journal} {\bibinfo
  {journal} {Rev. Mod. Phys.}\ }\textbf {\bibinfo {volume} {80}},\ \bibinfo
  {pages} {633} (\bibinfo {year} {2008})}\BibitemShut {NoStop}%
\bibitem [{\citenamefont {Pohl}\ \emph {et~al.}(2010)\citenamefont {Pohl} \emph
  {et~al.}}]{Pohl:2010zza}%
  \BibitemOpen
  \bibfield  {author} {\bibinfo {author} {\bibfnamefont {Randolf}\ \bibnamefont
  {Pohl}} \emph {et~al.},\ }\bibfield  {title} {\enquote {\bibinfo {title}
  {{The size of the proton}},}\ }\href {\doibase 10.1038/nature09250}
  {\bibfield  {journal} {\bibinfo  {journal} {Nature}\ }\textbf {\bibinfo
  {volume} {466}},\ \bibinfo {pages} {213} (\bibinfo {year}
  {2010})}\BibitemShut {NoStop}%
\bibitem [{\citenamefont {Antognini}\ \emph {et~al.}(2013)\citenamefont
  {Antognini} \emph {et~al.}}]{Antognini:1900ns}%
  \BibitemOpen
  \bibfield  {author} {\bibinfo {author} {\bibfnamefont {Aldo}\ \bibnamefont
  {Antognini}} \emph {et~al.},\ }\bibfield  {title} {\enquote {\bibinfo {title}
  {{Proton Structure from the Measurement of $2S-2P$ Transition Frequencies of
  Muonic Hydrogen}},}\ }\href {\doibase 10.1126/science.1230016} {\bibfield
  {journal} {\bibinfo  {journal} {Science}\ }\textbf {\bibinfo {volume}
  {339}},\ \bibinfo {pages} {417} (\bibinfo {year} {2013})}\BibitemShut
  {NoStop}%
\bibitem [{\citenamefont {Chang}\ \emph {et~al.}(2008)\citenamefont {Chang}
  \emph {et~al.}}]{Chang:2008aa}%
  \BibitemOpen
  \bibfield  {author} {\bibinfo {author} {\bibfnamefont {J.}~\bibnamefont
  {Chang}} \emph {et~al.},\ }\bibfield  {title} {\enquote {\bibinfo {title}
  {{An excess of cosmic ray electrons at energies of 300-800 GeV}},}\ }\href
  {\doibase 10.1038/nature07477} {\bibfield  {journal} {\bibinfo  {journal}
  {Nature}\ }\textbf {\bibinfo {volume} {456}},\ \bibinfo {pages} {362}
  (\bibinfo {year} {2008})}\BibitemShut {NoStop}%
\bibitem [{\citenamefont {Adriani}\ \emph {et~al.}(2009)\citenamefont {Adriani}
  \emph {et~al.}}]{Adriani:2008zr}%
  \BibitemOpen
  \bibfield  {author} {\bibinfo {author} {\bibfnamefont {Oscar}\ \bibnamefont
  {Adriani}} \emph {et~al.} (\bibinfo {collaboration} {PAMELA Collaboration}),\
  }\bibfield  {title} {\enquote {\bibinfo {title} {{An anomalous positron
  abundance in cosmic rays with energies 1.5-100 GeV}},}\ }\href {\doibase
  10.1038/nature07942} {\bibfield  {journal} {\bibinfo  {journal} {Nature}\
  }\textbf {\bibinfo {volume} {458}},\ \bibinfo {pages} {607} (\bibinfo {year}
  {2009})}\BibitemShut {NoStop}%
\bibitem [{\citenamefont {Aguilar}\ \emph {et~al.}(2013)\citenamefont {Aguilar}
  \emph {et~al.}}]{Aguilar:2013qda}%
  \BibitemOpen
  \bibfield  {author} {\bibinfo {author} {\bibfnamefont {M.}~\bibnamefont
  {Aguilar}} \emph {et~al.} (\bibinfo {collaboration} {AMS Collaboration}),\
  }\bibfield  {title} {\enquote {\bibinfo {title} {{First Result from the Alpha
  Magnetic Spectrometer on the International Space Station: Precision
  Measurement of the Positron Fraction in Primary Cosmic Rays of 0.5--350
  GeV}},}\ }\href {\doibase 10.1103/PhysRevLett.110.141102} {\bibfield
  {journal} {\bibinfo  {journal} {Phys. Rev. Lett.}\ }\textbf {\bibinfo
  {volume} {110}},\ \bibinfo {pages} {141102} (\bibinfo {year}
  {2013})}\BibitemShut {NoStop}%
\bibitem [{\citenamefont {Arkani-Hamed}\ \emph {et~al.}(2009)\citenamefont
  {Arkani-Hamed}, \citenamefont {Finkbeiner}, \citenamefont {Slatyer},\ and\
  \citenamefont {Weiner}}]{ArkaniHamed:2008qn}%
  \BibitemOpen
  \bibfield  {author} {\bibinfo {author} {\bibfnamefont {Nima}\ \bibnamefont
  {Arkani-Hamed}}, \bibinfo {author} {\bibfnamefont {Douglas~P.}\ \bibnamefont
  {Finkbeiner}}, \bibinfo {author} {\bibfnamefont {Tracy~R.}\ \bibnamefont
  {Slatyer}}, \ and\ \bibinfo {author} {\bibfnamefont {Neal}\ \bibnamefont
  {Weiner}},\ }\bibfield  {title} {\enquote {\bibinfo {title} {{A Theory of
  Dark Matter}},}\ }\href {\doibase 10.1103/PhysRevD.79.015014} {\bibfield
  {journal} {\bibinfo  {journal} {Phys. Rev. D}\ }\textbf {\bibinfo {volume}
  {79}},\ \bibinfo {pages} {015014} (\bibinfo {year} {2009})}\BibitemShut
  {NoStop}%
\bibitem [{\citenamefont {Tucker-Smith}\ and\ \citenamefont
  {Yavin}(2011)}]{TuckerSmith:2010ra}%
  \BibitemOpen
  \bibfield  {author} {\bibinfo {author} {\bibfnamefont {David}\ \bibnamefont
  {Tucker-Smith}}\ and\ \bibinfo {author} {\bibfnamefont {Itay}\ \bibnamefont
  {Yavin}},\ }\bibfield  {title} {\enquote {\bibinfo {title} {{Muonic hydrogen
  and MeV forces}},}\ }\href {\doibase 10.1103/PhysRevD.83.101702} {\bibfield
  {journal} {\bibinfo  {journal} {Phys. Rev. D}\ }\textbf {\bibinfo {volume}
  {83}},\ \bibinfo {pages} {101702} (\bibinfo {year} {2011})}\BibitemShut
  {NoStop}%
\bibitem [{\citenamefont {Jaeckel}(2012)}]{Jaeckel:2013ija}%
  \BibitemOpen
  \bibfield  {author} {\bibinfo {author} {\bibfnamefont {J.}~\bibnamefont
  {Jaeckel}},\ }\bibfield  {title} {\enquote {\bibinfo {title} {{A force beyond
  the Standard Model- Status of the quest for hidden photons}},}\ }\href@noop
  {} {\bibfield  {journal} {\bibinfo  {journal} {Frascati Phys.~Ser.}\ }\textbf
  {\bibinfo {volume} {56}},\ \bibinfo {pages} {172} (\bibinfo {year}
  {2012})}\BibitemShut {NoStop}%
\bibitem [{\citenamefont {Essig}\ \emph {et~al.}()\citenamefont {Essig} \emph
  {et~al.}}]{Essig:2013lka}%
  \BibitemOpen
  \bibfield  {author} {\bibinfo {author} {\bibfnamefont {Rouven}\ \bibnamefont
  {Essig}} \emph {et~al.},\ }\href@noop {} {\enquote {\bibinfo {title} {{Dark
  Sectors and New, Light, Weakly-Coupled Particles}},}\ }\bibinfo {note}
  {ArXiv:1311.0029}\BibitemShut {NoStop}%
\bibitem [{\citenamefont {Kroll}\ and\ \citenamefont
  {Wada}(1955)}]{Kroll:1955zu}%
  \BibitemOpen
  \bibfield  {author} {\bibinfo {author} {\bibfnamefont {Norman~M.}\
  \bibnamefont {Kroll}}\ and\ \bibinfo {author} {\bibfnamefont {Walter}\
  \bibnamefont {Wada}},\ }\bibfield  {title} {\enquote {\bibinfo {title}
  {{Internal pair production associated with the emission of high-energy gamma
  rays}},}\ }\href {\doibase 10.1103/PhysRev.98.1355} {\bibfield  {journal}
  {\bibinfo  {journal} {Phys.~Rev.}\ }\textbf {\bibinfo {volume} {98}},\
  \bibinfo {pages} {1355} (\bibinfo {year} {1955})}\BibitemShut {NoStop}%
\bibitem [{\citenamefont {Dzhelyadin}\ \emph {et~al.}(1980)\citenamefont
  {Dzhelyadin} \emph {et~al.}}]{Dzhelyadin:1980kh}%
  \BibitemOpen
  \bibfield  {author} {\bibinfo {author} {\bibfnamefont {R.~I.}\ \bibnamefont
  {Dzhelyadin}} \emph {et~al.} (\bibinfo {collaboration} {SERPUKHOV-134
  Collaboration}),\ }\bibfield  {title} {\enquote {\bibinfo {title}
  {{Investigation of $\eta$ Meson Electromagnetic Structure in $\eta \to \mu^+
  \mu^- \gamma$ Decay}},}\ }\href {\doibase 10.1016/0370-2693(80)90937-5}
  {\bibfield  {journal} {\bibinfo  {journal} {Phys. Lett. B}\ }\textbf
  {\bibinfo {volume} {94}},\ \bibinfo {pages} {548} (\bibinfo {year}
  {1980})}\BibitemShut {NoStop}%
\bibitem [{\citenamefont {Adare}\ \emph
  {et~al.}(2013{\natexlab{a}})\citenamefont {Adare} \emph
  {et~al.}}]{Adare:2012vn}%
  \BibitemOpen
  \bibfield  {author} {\bibinfo {author} {\bibfnamefont {A.}~\bibnamefont
  {Adare}} \emph {et~al.} (\bibinfo {collaboration} {PHENIX Collaboration}),\
  }\bibfield  {title} {\enquote {\bibinfo {title} {{Direct photon production in
  $d+$Au collisions at $\sqrt{s_{NN}}=200$ GeV}},}\ }\href {\doibase
  10.1103/PhysRevC.87.054907} {\bibfield  {journal} {\bibinfo  {journal} {Phys.
  Rev. C}\ }\textbf {\bibinfo {volume} {87}},\ \bibinfo {pages} {054907}
  (\bibinfo {year} {2013}{\natexlab{a}})}\BibitemShut {NoStop}%
\bibitem [{\citenamefont {Anderson}\ \emph {et~al.}(2011)\citenamefont
  {Anderson} \emph {et~al.}}]{Anderson:2011jw}%
  \BibitemOpen
  \bibfield  {author} {\bibinfo {author} {\bibfnamefont {W.}~\bibnamefont
  {Anderson}} \emph {et~al.} (\bibinfo {collaboration} {PHENIX
  Collaboration}),\ }\bibfield  {title} {\enquote {\bibinfo {title} {{Design,
  Construction, Operation and Performance of a Hadron Blind Detector for the
  PHENIX Experiment}},}\ }\href {\doibase 10.1016/j.nima.2011.04.015}
  {\bibfield  {journal} {\bibinfo  {journal} {Nucl. Instrum. Methods Phys.
  Res., Sect. A}\ }\textbf {\bibinfo {volume} {646}},\ \bibinfo {pages} {35}
  (\bibinfo {year} {2011})}\BibitemShut {NoStop}%
\bibitem [{\citenamefont {Adcox}\ \emph {et~al.}(2003)\citenamefont {Adcox}
  \emph {et~al.}}]{Adcox:2003zm}%
  \BibitemOpen
  \bibfield  {author} {\bibinfo {author} {\bibfnamefont {K.}~\bibnamefont
  {Adcox}} \emph {et~al.} (\bibinfo {collaboration} {PHENIX Collaboration}),\
  }\bibfield  {title} {\enquote {\bibinfo {title} {{PHENIX detector
  overview}},}\ }\href {\doibase 10.1016/S0168-9002(02)01950-2} {\bibfield
  {journal} {\bibinfo  {journal} {Nucl. Instrum. Methods Phys. Res., Sect. A}\
  }\textbf {\bibinfo {volume} {499}},\ \bibinfo {pages} {469} (\bibinfo {year}
  {2003})}\BibitemShut {NoStop}%
\bibitem [{\citenamefont {Adare}\ \emph {et~al.}(2010)\citenamefont {Adare}
  \emph {et~al.}}]{Adare:2009qk}%
  \BibitemOpen
  \bibfield  {author} {\bibinfo {author} {\bibfnamefont {A.}~\bibnamefont
  {Adare}} \emph {et~al.} (\bibinfo {collaboration} {PHENIX Collaboration}),\
  }\bibfield  {title} {\enquote {\bibinfo {title} {{Detailed measurement of the
  $e^+e^-$ pair continuum in $p+p$ and Au+Au collisions at $\sqrt{s_{NN}}$ =
  200 GeV and implications for direct photon production}},}\ }\href {\doibase
  10.1103/PhysRevC.81.034911} {\bibfield  {journal} {\bibinfo  {journal} {Phys.
  Rev. C}\ }\textbf {\bibinfo {volume} {81}},\ \bibinfo {pages} {034911}
  (\bibinfo {year} {2010})}\BibitemShut {NoStop}%
\bibitem [{\citenamefont {Adare}\ \emph
  {et~al.}(2013{\natexlab{b}})\citenamefont {Adare} \emph
  {et~al.}}]{Adare:2012vv}%
  \BibitemOpen
  \bibfield  {author} {\bibinfo {author} {\bibfnamefont {A.}~\bibnamefont
  {Adare}} \emph {et~al.} (\bibinfo {collaboration} {PHENIX Collaboration}),\
  }\bibfield  {title} {\enquote {\bibinfo {title} {{Double Spin Asymmetry of
  Electrons from Heavy Flavor Decays in $p+p$ Collisions at $\sqrt{s}=200$
  GeV}},}\ }\href {\doibase 10.1103/PhysRevD.87.012011} {\bibfield  {journal}
  {\bibinfo  {journal} {Phys. Rev. D}\ }\textbf {\bibinfo {volume} {87}},\
  \bibinfo {pages} {012011} (\bibinfo {year} {2013}{\natexlab{b}})}\BibitemShut
  {NoStop}%
\bibitem [{\citenamefont {Adare}\ \emph {et~al.}()\citenamefont {Adare} \emph
  {et~al.}}]{PPG162:arXiv}%
  \BibitemOpen
  \bibfield  {author} {\bibinfo {author} {\bibfnamefont {A.}~\bibnamefont
  {Adare}} \emph {et~al.} (\bibinfo {collaboration} {PHENIX Collaboration}),\
  }\href@noop {} {\enquote {\bibinfo {title} {{Centrality dependence of
  low-momentum direct-photon production in Au+Au collisions at $\sqrt{s_{NN}}$
  = 200 GeV}},}\ }\bibinfo {note} {ArXiv:1405.3940}\BibitemShut {NoStop}%
\bibitem [{\citenamefont {Read}(2002)}]{Read:2002hq}%
  \BibitemOpen
  \bibfield  {author} {\bibinfo {author} {\bibfnamefont {Alexander~L.}\
  \bibnamefont {Read}},\ }\bibfield  {title} {\enquote {\bibinfo {title}
  {{Presentation of search results: The CL(s) technique}},}\ }\href {\doibase
  10.1088/0954-3899/28/10/313} {\bibfield  {journal} {\bibinfo  {journal} {J.
  Phys. G}\ }\textbf {\bibinfo {volume} {28}},\ \bibinfo {pages} {2693}
  (\bibinfo {year} {2002})}\BibitemShut {NoStop}%
\bibitem [{\citenamefont {Gross}\ and\ \citenamefont
  {Vitells}(2010)}]{lookelse}%
  \BibitemOpen
  \bibfield  {author} {\bibinfo {author} {\bibfnamefont {E.}~\bibnamefont
  {Gross}}\ and\ \bibinfo {author} {\bibfnamefont {O.}~\bibnamefont
  {Vitells}},\ }\bibfield  {title} {\enquote {\bibinfo {title} {{Trial factors
  for the look elsewhere effect in high energy physics}},}\ }\href@noop {}
  {\bibfield  {journal} {\bibinfo  {journal} {Eur. Phys. J. C}\ }\textbf
  {\bibinfo {volume} {70}},\ \bibinfo {pages} {525} (\bibinfo {year}
  {2010})}\BibitemShut {NoStop}%
\bibitem [{\citenamefont {Adlarson}\ \emph {et~al.}(2013)\citenamefont
  {Adlarson} \emph {et~al.}}]{Adlarson:2013eza}%
  \BibitemOpen
  \bibfield  {author} {\bibinfo {author} {\bibfnamefont {P.}~\bibnamefont
  {Adlarson}} \emph {et~al.} (\bibinfo {collaboration} {WASA-at-COSY
  Collaboration}),\ }\bibfield  {title} {\enquote {\bibinfo {title} {{Search
  for a dark photon in the $\pi^0 \to e^+e^-\gamma$ decay}},}\ }\href {\doibase
  10.1016/j.physletb.2013.08.055} {\bibfield  {journal} {\bibinfo  {journal}
  {Phys. Lett. B}\ }\textbf {\bibinfo {volume} {726}},\ \bibinfo {pages} {187}
  (\bibinfo {year} {2013})}\BibitemShut {NoStop}%
\bibitem [{\citenamefont {Agakishiev}\ \emph {et~al.}(2014)\citenamefont
  {Agakishiev} \emph {et~al.}}]{Agakishiev:2013fwl}%
  \BibitemOpen
  \bibfield  {author} {\bibinfo {author} {\bibfnamefont {G.}~\bibnamefont
  {Agakishiev}} \emph {et~al.} (\bibinfo {collaboration} {HADES
  Collaboration}),\ }\bibfield  {title} {\enquote {\bibinfo {title} {{Searching
  a Dark Photon with HADES}},}\ }\href {\doibase
  10.1016/j.physletb.2014.02.035} {\bibfield  {journal} {\bibinfo  {journal}
  {Phys. Lett. B}\ }\textbf {\bibinfo {volume} {731}},\ \bibinfo {pages} {265}
  (\bibinfo {year} {2014})}\BibitemShut {NoStop}%
\bibitem [{\citenamefont {Babusci}\ \emph {et~al.}(2013)\citenamefont {Babusci}
  \emph {et~al.}}]{Babusci:2012cr}%
  \BibitemOpen
  \bibfield  {author} {\bibinfo {author} {\bibfnamefont {D.}~\bibnamefont
  {Babusci}} \emph {et~al.} (\bibinfo {collaboration} {KLOE-2 Collaboration}),\
  }\bibfield  {title} {\enquote {\bibinfo {title} {{Limit on the production of
  a light vector gauge boson in phi meson decays with the KLOE detector}},}\
  }\href {\doibase 10.1016/j.physletb.2013.01.067} {\bibfield  {journal}
  {\bibinfo  {journal} {Phys. Lett. B}\ }\textbf {\bibinfo {volume} {720}},\
  \bibinfo {pages} {111} (\bibinfo {year} {2013})}\BibitemShut {NoStop}%
\bibitem [{\citenamefont {Merkel}\ \emph {et~al.}(2014)\citenamefont {Merkel}
  \emph {et~al.}}]{Merkel:2014avp}%
  \BibitemOpen
  \bibfield  {author} {\bibinfo {author} {\bibfnamefont {H.}~\bibnamefont
  {Merkel}} \emph {et~al.} (\bibinfo {collaboration} {MAMI Collaboration}),\
  }\bibfield  {title} {\enquote {\bibinfo {title} {{Search for light massive
  gauge bosons as an explanation of the $(g-2)_\mu$ anomaly at MAMI}},}\ }\href
  {\doibase 10.1103/PhysRevLett.112.221802} {\bibfield  {journal} {\bibinfo
  {journal} {Phys. Rev. Lett.}\ }\textbf {\bibinfo {volume} {112}},\ \bibinfo
  {pages} {221802} (\bibinfo {year} {2014})}\BibitemShut {NoStop}%
\bibitem [{\citenamefont {Lees}\ \emph {et~al.}(2014)\citenamefont {Lees} \emph
  {et~al.}}]{Lees:2014xha}%
  \BibitemOpen
  \bibfield  {author} {\bibinfo {author} {\bibfnamefont {J.P.}\ \bibnamefont
  {Lees}} \emph {et~al.} (\bibinfo {collaboration} {{\sc BaBar}
  Collaboration}),\ }\bibfield  {title} {\enquote {\bibinfo {title} {{Search
  for a Dark Photon in $e^+e^-$ Collisions at BaBar}},}\ }\href {\doibase
  10.1103/PhysRevLett.113.201801} {\bibfield  {journal} {\bibinfo  {journal}
  {Phys. Rev. Lett.}\ }\textbf {\bibinfo {volume} {113}},\ \bibinfo {pages}
  {201801} (\bibinfo {year} {2014})}\BibitemShut {NoStop}%
\bibitem [{\citenamefont {Davoudiasl}\ \emph {et~al.}(2014)\citenamefont
  {Davoudiasl}, \citenamefont {Lee},\ and\ \citenamefont
  {Marciano}}]{Davoudiasl:2014kua}%
  \BibitemOpen
  \bibfield  {author} {\bibinfo {author} {\bibfnamefont {Hooman}\ \bibnamefont
  {Davoudiasl}}, \bibinfo {author} {\bibfnamefont {Hye-Sung}\ \bibnamefont
  {Lee}}, \ and\ \bibinfo {author} {\bibfnamefont {William~J.}\ \bibnamefont
  {Marciano}},\ }\bibfield  {title} {\enquote {\bibinfo {title} {{Muon g-2,
  Rare Kaon Decays, and Parity Violation from Dark Bosons}},}\ }\href {\doibase
  10.1103/PhysRevD.89.095006} {\bibfield  {journal} {\bibinfo  {journal} {Phys.
  Rev. D}\ }\textbf {\bibinfo {volume} {89}},\ \bibinfo {pages} {095006}
  (\bibinfo {year} {2014})}\BibitemShut {NoStop}%
\bibitem [{\citenamefont {Maltoni}\ and\ \citenamefont
  {Schwetz}(2003)}]{Maltoni:2003cu}%
  \BibitemOpen
  \bibfield  {author} {\bibinfo {author} {\bibfnamefont {M.}~\bibnamefont
  {Maltoni}}\ and\ \bibinfo {author} {\bibfnamefont {T.}~\bibnamefont
  {Schwetz}},\ }\bibfield  {title} {\enquote {\bibinfo {title} {{Testing the
  statistical compatibility of independent data sets}},}\ }\href {\doibase
  10.1103/PhysRevD.68.033020} {\bibfield  {journal} {\bibinfo  {journal} {Phys.
  Rev. D}\ }\textbf {\bibinfo {volume} {68}},\ \bibinfo {pages} {033020}
  (\bibinfo {year} {2003})}\BibitemShut {NoStop}%
\bibitem [{\citenamefont {Bjorken}\ \emph {et~al.}(2009)\citenamefont
  {Bjorken}, \citenamefont {Essig}, \citenamefont {Schuster},\ and\
  \citenamefont {Toro}}]{Bjorken:2009mm}%
  \BibitemOpen
  \bibfield  {author} {\bibinfo {author} {\bibfnamefont {James~D.}\
  \bibnamefont {Bjorken}}, \bibinfo {author} {\bibfnamefont {Rouven}\
  \bibnamefont {Essig}}, \bibinfo {author} {\bibfnamefont {Philip}\
  \bibnamefont {Schuster}}, \ and\ \bibinfo {author} {\bibfnamefont {Natalia}\
  \bibnamefont {Toro}},\ }\bibfield  {title} {\enquote {\bibinfo {title} {{New
  Fixed-Target Experiments to Search for Dark Gauge Forces}},}\ }\href
  {\doibase 10.1103/PhysRevD.80.075018} {\bibfield  {journal} {\bibinfo
  {journal} {Phys. Rev. D}\ }\textbf {\bibinfo {volume} {80}},\ \bibinfo
  {pages} {075018} (\bibinfo {year} {2009})}\BibitemShut {NoStop}%
\end{thebibliography}

%
 
\end{document}